\documentclass[12pt]{iopart}
\usepackage{latexsym}
\usepackage{amscd}
\usepackage[dvips]{color}

\begin{document}
\definecolor{r}{rgb}{1,0,0}
\definecolor{b}{rgb}{0,0,1}
\definecolor{g}{cmyk}{0,0,1,0}
\jl{1}

\jl{1}
 \def\lambdabar{\protect\@lambdabar}
\def\@lambdabar{%
\relax
\bgroup
\def\@tempa{\hbox{\raise.73\ht0
\hbox to0pt{\kern.25\wd0\vrule width.5\wd0
height.1pt depth.1pt\hss}\box0}}%
\mathchoice{\setbox0\hbox{$\displaystyle\lambda$}\@tempa}%
{\setbox0\hbox{$\textstyle\lambda$}\@tempa}%
{\setbox0\hbox{$\scriptstyle\lambda$}\@tempa}%
{\setbox0\hbox{$\scriptscriptstyle\lambda$}\@tempa}%
\egroup
}

\def\bbox#1{%
\relax\ifmmode
\mathchoice
{{\hbox{\boldmath$\displaystyle#1$}}}%
{{\hbox{\boldmath$\textstyle#1$}}}%
{{\hbox{\boldmath$\scriptstyle#1$}}}%
{{\hbox{\boldmath$\scriptscriptstyle#1$}}}%
\else
\mbox{#1}%
\fi
}
\newcommand{\muv}{\bbox{\mu}}
\newcommand{\mc}{{\mathcal M}}
\newcommand{\pc}{{\mathcal P}}
\newcommand{\mct}{\bbox{\mathcal M}}
\newcommand{\pct}{\bbox {\mathcal P}}
\newcommand{\fsf}{{\sf F}}
\newcommand{\fsft}{\bbox{{\sf F}}}
 \newcommand{\mv}{\bbox{m}}
\newcommand{\pv}{\bbox{p}}
\newcommand{\tv}{\bbox{t}}
\def\msf{\hbox{{\sf M}}}
\def\msft{\bbox{{\sf M}}}
\def\psf{\hbox{{\sf P}}}
\def\psft{\bbox{{\sf P}}}
\def\Nsf{\hbox{{\sf N}}}
\def\Nsft{\bbox{{\sf N}}}
\def\Tsf{\hbox{{\sf T}}}
\def\Tsft{\bbox{{\sf T}}}
\def\Asf{\hbox{{\sf A}}}
\def\Asft{\bbox{{\sf A}}}
\def\Bsf{\hbox{{\sf B}}}
\def\Bsft{\bbox{{\sf B}}}
\def\Lsf{\hbox{{\sf L}}}
\def\Lsft{\bbox{{\sf L}}}
\def\Ssf{\hbox{{\sf S}}}
\def\Ssft{\bbox{{\sf S}}}
\def\Mtens{\bi{M}}
\def\msfsim{\bbox{{\sf M}}_{\scriptstyle\rm(sym)}}
\newcommand{\mcsim}{ {\sf M}_{ {\scriptstyle \rm {(sym)} } i_1\dots i_n}}
\newcommand{\mcs}{ {\sf M}_{ {\scriptstyle \rm {(sym)} } i_1i_2i_3}}

\newcommand{\beqan}{\begin{eqnarray*}}
\newcommand{\eeqan}{\end{eqnarray*}}
\newcommand{\beqa}{\begin{eqnarray}}
\newcommand{\eeqa}{\end{eqnarray}}

 \newcommand{\suml}{\sum\limits}
 \newcommand{\sumd}{\suml_{\mathcal D}}
\newcommand{\intl}{\int\limits}
\newcommand{\rvec}{\bbox{r}}
\newcommand{\xivec}{\bbox{\xi}}
\newcommand{\Avec}{\bbox{A}}
\newcommand{\Rvec}{\bbox{R}}
\newcommand{\Evec}{\bbox{E}}
\newcommand{\Bvec}{\bbox{B}}
\newcommand{\Svec}{\bbox{S}}
\newcommand{\avec}{\bbox{a}}
\newcommand{\nablav}{\bbox{\nabla}}
\newcommand{\nuvec}{\bbox{\nu}}
\newcommand{\bvec}{\bbox{\beta}}
\newcommand{\vvec}{\bbox{v}}
\newcommand{\jvec}{\bbox{J}}
\newcommand{\nvec}{\bbox{n}}
\newcommand{\pvec}{\bbox{p}}
\newcommand{\mvec}{\bbox{m}}
\newcommand{\evec}{\bbox{e}}
\newcommand{\eps}{\varepsilon}
\newcommand{\la}{\lambda}
\newcommand{\rad}{\mbox{\footnotesize rad}}
\newcommand{\scr}{\scriptstyle}
\newcommand{\latens}{\bbox{\sf{\Lambda}}}
\newcommand{\lasf}{{\sf{\Lambda}}}
\newcommand{\pitens}{\sf{\Pi}}
\newcommand{\cm}{{\cal M}}
\newcommand{\cp}{{\cal P}}
\newcommand{\beq}{\begin{equation}} 
\newcommand{\eeq}{\end{equation}}
\newcommand{\ptens}{\bbox{\sf{P}}}
\newcommand{\Ptens}{\bbox{P}}
\newcommand{\Ttens}{\bbox{\sf{T}}}
\newcommand{\Ntens}{\bbox{\sf{N}}}
\newcommand{\Ncal}{\bbox{{\cal N}}}
\newcommand{\Atens}{\bbox{\sf{A}}}
\newcommand{\Btens}{\bbox{\sf{B}}}
\newcommand{\dom}{\mathcal{D}}
\newcommand{\al}{\alpha}
\newcommand{\sym}{\scriptstyle \rm{(sym)}}
\newcommand{\Tcal}{\bbox{{\mathcal T}}}
\newcommand{\Nmc}{{\mathcal N}}
\renewcommand{\d}{\partial}
\def\rmi{{\rm i}}
\def\rme{\hbox{\rm e}}
\def\rmd{\hbox{\rm d}}
\newcommand{\ct}{\mbox{\Huge{.}}}
\newcommand{\Laop}{\bbox{\Lambda}}
\newcommand{\Ssfs}{{\scriptstyle \Ssft^{(n)}}}
\newcommand{\Lsfs}{{\scriptstyle \Lsft^{(n)}}}
\newcommand{\psfr}{\widetilde{\psf}}
\newcommand{\msfr}{\widetilde{\msf}}
\newcommand{\msftr}{\widetilde{\msft}}
\newcommand{\psftr}{\widetilde{\psft}}
\newcommand{\pvr}{\widetilde{\pvec}}
\newcommand{\mvr}{\widetilde{\mvec}}
\newcommand{\qdot}{\stackrel{\cdot\cdot\cdot\cdot}}
\newcommand{\bsy}{\hbox}
\newcommand{\ointl}{\oint\limits}
\newcommand{\pisf}{{\sf \Pi}}
\def\Nsf{{\sf N}}
\newcommand{\gamsf}{{\sf \Gamma}}
\newcommand{\gamsft}{\bsy{\sf \Gamma}}
\newcommand{\ab}{\v{a}} 
\newcommand{\ai}{\^{a}} 
\newcommand{\ib}{\^{\i}} 
\newcommand{\tb}{\c{t}} 
\newcommand{\st}{\c{s}}
\newcommand{\Ab}{\v{A}} 
\newcommand{\Ai}{\^{A}} 
\newcommand{\Ib}{\^{I}} 
\newcommand{\Tb}{\c{T}}
\newcommand{\St}{\c{S}}
\newcommand{\mlrt}{\stackrel{\leftrightarrow}{\msft}}
\newcommand{\mlr}{\stackrel{\leftrightarrow}{\msf}}
\title{Singular behaviour of the electromagnetic field (the static case revisited) }
\author{C.\ Vrejoiu\footnote{E-mail :  vrejoiu@fizica.unibuc.ro} and R.\ Zus \footnote{E-mail: roxana.zus@fizica.unibuc.ro}}
\address{University of Bucharest, Department of Physics,  \\
PO Box MG - 11, Bucharest-Magurele, RO - 077125,
 Romania }

\begin{abstract}
 The singularities of the electromagnetic field  are derived to include all the point-like multipoles representing an electric charge and current distribution. Partial results obtained in a previous paper \cite{cvrz-arx} are completed to represent accurately all the terms included in these singularities.\end{abstract}
\section{Introduction}\label{intro}
In the cases of electrostatic and magnetostatic fields of point-like dipoles, one has the well-known procedure of introducing  Dirac $\delta$-function terms for obtaining correct expressions of the electric and magnetic fields defined on the entire space.  The corresponding field expressions take the following form \cite{Jackson}: 
\beqa\label{1}
\Evec_{\pvec}(\rvec)=-\frac{1}{3\eps_0}\,\pvec\,\delta(\rvec)+\frac{1}{4\pi\eps_0}\frac{3(\nuvec\cdot\pvec)\nuvec-\pvec}{r^3}=-\frac{1}{3\eps_0}\,\pvec\,\delta(\rvec) + \big(\Evec\big)_{r\ne 0} ,
\eeqa
where $\nuvec=\rvec/r$, and
\beqa\label{2}
\Bvec_{\mvec}(\rvec)= \frac{2\mu_0}{3}\,\mvec\,\delta(\rvec)+\frac{\mu_0}{4\pi}\frac{3(\nuvec\cdot\mvec)\nuvec-\mvec}{r^3}=
\frac{2\mu_0}{3}\,\mvec\,\delta(\rvec)+ \big(\Bvec\big)_{r\ne 0} .
\eeqa
In these equations, by $(\dots)_{r\ne 0}$ we understand an expression in which the derivatives are calculated supposing $r\ne 0$, representing some well-known expressions of the fields.
The expressions from equations \eref{1} and \eref{2} are introduced in Ref.\  \cite{Jackson} as  conditions of compatibility with the average value of the  electric or magnetic field  over a spherical domain containing all the charges or currents inside. Another procedure for introducing equations \eref{1} and \eref{2} is based on an extension of the derivative $\d_i\d_j/(1/r)$ to the entire space \cite{Frahm}:
\beqa\label{3}
\d_i\d_j\frac{1}{r}=\,-\frac{4\pi}{3}\,\delta_{ij}\,\delta(\rvec)\,+\,\frac{3\,\nu_i\nu_j\,-\,\delta_{ij}}{r^3}.
\eeqa
A more pedagogical and suitable approach for understanding the origin of the difference between  the electric and magnetic cases  is done in Ref.\  \cite{Leung06}.  Refs.\  \cite{Werner} and \cite{Leung07}  contain  generalizations of the equations \eref{1} and \eref{2} to the dynamic case for  oscillating electric and magnetic dipoles.
\par The objective of the present paper is to establish the singularities of the electromagnetic field associated to a system of electric charges and currents assimilated with a point-like multipolar system. These singularities are established for an arbitrary multipolar order in  the static  case.
\par In section 2, the procedure of separating the $\delta$-form singularities of the multiple partial derivatives of  $r^{-1}$ is described. In  sections 3 and 4, we separate the $\delta$-form singularities of the fields corresponding to the point-like equivalent multipole distributions, based on the multipole expansions of the fields associated to an electric charge or current distribution confined in a domain  $\dom$.
\par We point out that the formalism presented in this paper has as mathematical basis the properties of the irreducible tensorial  representations of the proper rotations group \cite{Damour}. For pedagogical  and larger accessibility reasons, we give an explicit calculation based on the properties of the tensor contractions such that the procedure has a simple  algebraic character. A counterpart of the procedure used in the present paper could be represented by the technique of the spherical function expansions and the cited issues can be a basis for such an approach. However, the use of the spherical coordinates can lead to the omission of the $\delta$-form singularity contributions, as argued in the conclusions.

\section{Some delta-function identities}\label{delta}

The treatment of some delta-function identities in Ref.  \cite{Frahm} can be easily generalized to obtain the necessary identities for higher order derivatives of $1/r$. In the points different from the origin $O$, the function $1/r$ is a solution of the Laplace equation: 
\beqa\label{4}
\Delta\frac{1}{r}=0,\;\;\;\;r\ne 0\ .
\eeqa
 It is well-known how one can extend $\Delta(1/r)$ as a distribution to the entire space, as solution of the Poisson equation:
 \beqa\label{5}
\Delta\frac{1}{r}=-4\pi\,\delta(\rvec)\ .
\eeqa 
Equation \eref{3} represents the next step of generalization and can be continued for arbitrary multiple partial derivatives of $1/r$ i.e. for $\d_{i_1}\dots \d_{i_n}\,(1/r)$ for arbitrary $n$. For $r\ne 0$, this multiple partial derivative is given by the formula
\beqa\label{6}
 \left(\d_{i_1}\dots \d_{i_n}\,\frac{1}{r}\right)_{r\ne 0}=\frac{1}{r^{n+1}}C^{(n)}_{i_1\dots i_n}\ .
 \eeqa
In this equation, the coefficient $C$ is fully symmetric with respect to $i_1\dots i_n$ and can be written as
\beqa\label{7}
C^{(n)}_{i_1\dots i_n}=\suml^{[\frac{n}{2}]}_{k}\,(-1)^k(2n-2k-1)!!\delta_{\{i_1i_2}\dots \delta_{i_{2k-1}i_{2k}}\nu_{2k+1}\dots \nu_{i_n\}}\ .
\eeqa
In the last equation, $[\beta]$ is the integer part of $\beta$ and by 
$A_{\{i_1\dots i_n\}}$ 
we understand the sum over all the  permutations of the symbols $i_q$ giving distinct terms.  From equation \eref{5}, it is obvious that we have to consider the derivatives of $1/r$ extended over the entire space as  distributions (generalized functions) which contain singular distributions ($\delta$-functions for example) as separate terms.
 \par Let be a function $F(\rvec)$ and suppose the existence of the integral of the product $F(\rvec)\,\phi(\rvec)$, with $\phi(\rvec)$ an arbitrary smooth function (a test function from the domain of the distributions),  
 on the spherical region $\dom_R$, with arbitrary radius $R$,  delimited by the spherical surface $\Sigma_R$ with the center  in $O$:
  \beqa\label{8}
 \int_{\dom_R}\rmd^3x\,F(\rvec)\,\phi(\rvec)=\lim_{\eps\to 0}\int_{\eps<r < R}\rmd^3x\,F(\rvec)\,\phi(\rvec)\ .
 \eeqa
 This integral can be expressed excluding from the domain $\dom_R$ a spherical domain of radius $\eps$ centered in $O$. Writing this last limit of integrals, we can interpret the function $F(\rvec)$ as a distribution  defined by
 \beqan
 \left\langle\,\left(F(\rvec)\right)_{r\ne 0},\;\phi(\rvec)\right\rangle=\lim_{\eps\to 0}\int_{\dom_R\setminus\dom_\eps}\rmd^3x\,(F(\rvec))_{r\ne 0}\,\phi(\rvec)\ .
 \eeqan
 This distribution can be extended such that its support includes the point $O$. A new term $\theta(\eps-r)\,F(\rvec)$ can be naturally  introduced by the identity
 \beqan
 F_\eps(\rvec)=\theta(\eps-r)\,F(\rvec,t)+\theta(r-\eps)F(\rvec),
 \eeqan
 associated with the extension of the integral to the entire domain $\dom_R$:
  \beqa\label{9}
 \left\langle\,F(\rvec),\,\phi(\rvec)\right\rangle=\lim_{\eps\to 0}\left[\int_{\dom_\eps}\rmd^3x\,F(\rvec)\,\phi(\rvec)+\int_{\dom_R\setminus\dom_\eps}\rmd^3x\,F(\rvec)\,\phi(\rvec)\right].
 \eeqa
Moreover, we suppose the existence of the integral \eref{8} for the partial derivatives of $F$.
  Let us consider the partial derivative $\d_iF(\rvec,t)$ and the problem of extending this function as a distribution. 
  The definition \eref{9} becomes:
  \beqa\label{10}
 \left\langle\,F(\rvec,t),\,\phi(\rvec)\right\rangle&=&\lim_{\eps\to 0}\left[\oint_{\Sigma_\eps}\rmd S\,\nu_i\,F(\rvec,t)\phi(\rvec)-\int_{\dom_\eps}\rmd^3 x\,F(\rvec,t)\,\d_i\phi(\rvec)\right.\nonumber\\
&&\,\ \ +\left.\int_{\dom_R\setminus \dom_\eps}\,\rmd^3x\,\d_iF(\rvec,t)\,\phi(\rvec)\right],
\eeqa
where $\Sigma_{\eps}$ is the sphere of radius $\eps$ centered in $O$  and the Gauss theorem was employed.\\
Let us apply this definition to the derivative $\d_i\d_j(1/r)$ and, for simplifying the notation, let 
\beqan
D_{i_1\dots i_n}(\rvec,t)=\d_{i_1}\dots \d_{i_n}\,\frac{1}{r},
\eeqan
such that we can write
\beqa\label{11}
\fl \left(D_{ij},\,\phi\right)= \lim_{\eps\to 0}\left[\oint_{\Sigma_\eps}\,\rmd S\,\nu_i\,\d_j\frac{1}{r}\,\phi(\rvec)- \int_{\dom_\eps}\,\rmd^3x\,\d_j\frac{1}{r}\,\d_i\phi(\rvec)\right.
 + \left. \int_{\dom_R\setminus\dom_\eps}\rmd^3x\,\big(\d_i\d_j\frac{1}{r}\big)\,\phi(\rvec)\right].              \eeqa
Since the last integral on the domain $\dom_R\setminus\dom_\eps$ represents the distributions associated with the $F$- expressions for $r\ne 0$ and, in  the case of the electromagnetic  field, they will be the well-known expressions of the multipole expansions, in the following, we consider only that part of $\left\langle D_{ij}\right\rangle$ containing singular distributions with point-like support i.e., actually,  the difference
\beqa\label{12}
\left\langle \left(D_{ij}\right)_{(0)},\;\phi\right\rangle&=&\left\langle D_{ij},\;\phi\right\rangle-\lim_{\eps\to 0}\int_{\dom_R\setminus\dom_\eps}\rmd^3x\,\big(\d_i\d_j\frac{1}{r}\big)\,\phi(\rvec)
\nonumber\\
&=&\lim_{\eps\to 0}\left[\oint_{\Sigma_\eps}\,\rmd S\,\nu_i\,\d_j\frac{1}{r}\,\phi(\rvec)- \int_{\dom_\eps}\,\rmd^3x\,\d_j\frac{1}{r}\,\d_i\phi(\rvec)\right]\ .
\eeqa
By $D_{(\rvec_0)}$ we denote a distribution having as support the point given by the vector $\rvec_0$. The surface integral,
\beqan
\lim_{\eps\to 0}\oint_{\Sigma_\eps}\,\rmd S\,\nu_i\,\d_j\frac{1}{r}\,\phi(\rvec)=-\lim_{\eps\to 0}\oint_{\Sigma_\eps}\,\rmd S\,\frac{1}{r^2}\nu_i\nu_j\phi(\rvec)\ , 
\eeqan
after inserting the Taylor series of the function $\phi(\rvec)$ and since on the sphere $r=\eps$, becomes  \cite{Frahm},
\beqa\label{13}
\lim_{\eps\to 0}\oint_{\Sigma_\eps}\,\rmd S\,\nu_i\,\d_j\frac{1}{r}\,\phi(\rvec) =-\lim_{\eps\to 0}\,\int\rmd\Omega(\nuvec)\,\nu_i\nu_j
 \left[\phi(0)+\eps\nu_k\left(\d_k\phi\right)_0+\dots\right]\ .
\eeqa
Let us introduce the angular average:
\beqa\label{14}
\langle g(\nuvec)\rangle=\frac{1}{4\pi}\int\,g(\nuvec)\,\rmd\Omega(\nuvec)\ .
\eeqa
Particularly, we have the well-known  formula \cite{Thorne}:
\beqa\label{15}
\langle\nu_{i_1}\dots\nu_{i_n}\rangle=\left\{\begin{array}{c}0,\;\;\;\;\;\;\;\;\;\;\;\;\;\;\;\;\;\;\;\;\;\;\;\;\;\;n=2k+1,\\
\frac{1}{(n+1)!!}\,\delta_{\{i_1i_2}\dots\delta_{i_{n-1}i_n\}},\;\;\;\;\;\;\;\;\;\;n=2k,\;\;\;\;\;k=0,1,\dots\end{array}\right. 
\eeqa
  Excepting the term containing  $\phi(0)$, all the terms in equation \eref{14} are proportional to positive powers of $\eps$ and, consequently, vanish with $\eps\to 0$, such that
 \beqan
\fl\;\;\;\;\;\;\; \lim_{\eps\to 0}\oint_{\Sigma_\eps}\,\rmd S\,\nu_i\,\d_j\frac{1}{r}\,\phi(\rvec)=-4\pi\left\langle\,\nu_i\nu_j\right\rangle\,\phi(0)
=-\frac{4\pi}{3}\delta_{ij}\phi(0)=-\frac{4\pi}{3}\delta_{ij}\left\langle\delta(\rvec),\;\phi(\rvec)\right\rangle\ .
 \eeqan
  Considering the second integral in the right-hand side of equation \eref{11}, we can write
  \beqan
\fl \lim_{\eps\to 0}\int_{\dom_\eps}\rmd^3x\,\,\d_j\frac{1}{r}\,\d_i\phi(\rvec)
=-\lim_{\eps\to 0}\int^\eps_0r^2\,\rmd    r\int\rmd\Omega(\nuvec)\,\nu_j\frac{1}{r^2}\big[(\d_i\phi)_0+r\nu_k(\d_i\d_k\phi)_0+\dots\big]=0\ ,
\eeqan
such that finally
\beqa\label{16}
\left(\d_i\d_j\frac{1}{r}\right)_{(0)}=-\frac{4\pi}{3}\delta_{ij}\,\delta(\rvec)\ ,
 \eeqa      
 i.e.   the delta-singularity from equation (3) of Ref.\  \cite{Frahm}.
\par  \par Let us consider the distribution $D_{ijk}$. Considering only the part having $O$ as support,
\beqa\label{17}
\fl \;\;\;\;\;\;\;\;\;\;\;\;\left(\left(D_{ijk}\right)_{(0)},\,\phi\right)
=\lim_{\eps\to 0}\left[\oint_{\Sigma_\eps}\rmd S\,\nu_i\,\d_j\d_k\frac{1}{r}\phi(\rvec)-
\int_{\dom_\eps}\rmd^3x\,\big(\d_j\d_k\frac{1}{r}\big)\,\d_i\phi(\rvec)\right].
\eeqa
 The surface integral becomes:
 \beqan
 \lim_{\eps\to 0}\oint_{\Sigma_\eps}\rmd S\,\nu_i\,\d_j\d_k\frac{1}{r}\,\phi(\rvec)&=&\lim_{\eps\to 0}\oint_{\Sigma_\eps}\rmd S\,\nu_i
 \frac{1}{r^3}(3\nu_j\nu_k-\delta_{jk})\phi(\rvec)
 \eeqan
 and, introducing the Taylor series for  $\phi(\rvec)$,
 \beqa\label{18}
 \fl \;\;\;\;\lim_{\eps\to 0}\oint_{\Sigma_\eps}\rmd S\,\nu_i\,\d_j\d_k\frac{1}{r}\,\phi(\rvec)=
 4\pi\,\lim_{\eps\to 0}\left\langle\,
 \frac{1}{\eps}(3\nu_i\nu_j\nu_k-\nu_{i}\delta_{jk}) 
\left[\phi(0)+\eps\nu_l\left(\d_l\phi\right)_0+\dots\right]\right\rangle\, ,
 \eeqa
and
\beqa\label{19}
\lim_{\eps\to 0}\oint_{\Sigma_\eps}\rmd S\,\nu_i\,\d_j\d_k\frac{1}{r}\,\phi(\rvec)\nonumber\\
=4\pi\left\langle3\nu_i\nu_j\nu_k\nu_l-\nu_i\nu_l\delta_{jk}\right\rangle\left(\d_l\phi\right)_0
=4\pi\left(\frac{1}{5}\delta_{\{ij}\delta_{kl\}}-\frac{1}{3}\delta_{il}\delta_{jk}\right)\,\left(\d_l\phi\right)_0\ .
\eeqa
 Concerning the integral on $\dom_\eps$ from equation \eref{17}, we have to observe that, beginning from this derivative order, there is a non-zero contribution for $\eps\to 0$ \cite{Frahm}. Indeed, introducing equation \eref{16} in equation \eref{17} and noticing that the term $(\d_j\d_k(f(\tau)/r))_{r\ne 0}$ gives a null contribution to the limit for $\eps\to 0$, we can write
\beqan
 -\lim_{\eps\to 0}\int_{\dom_\eps}\rmd^3x\,\d_j\d_k\frac{1}{r}\,\,\d_i\phi(\rvec)=\frac{4\pi}{3}\int_{\dom_\eps}\rmd^3x\,\delta_{jk}\delta(\rvec)\d_i\phi(\rvec)=\frac{4\pi}{3}\,\delta_{jk}(\d_i\phi)_0\ .
\eeqan
Finally, equations \eref{17}, \eref{18} and \eref{19} give 
\beqa\label{20}
(D_{ijk})_{(0)}=-\frac{4\pi}{5}\,\delta_{\{ij}\d_{k\}}\delta(\rvec)\ ,
\eeqa
i.e.  the delta-singularity from equation (4) of Ref.\  \cite{Frahm}.
\par Obviously, this procedure becomes very complicated for higher order derivatives. Fortunately, for the electromagnetic field, some invariance properties allow a considerable simplification of such calculations.

\section{Singularities of the electrostatic field}\label{e-static}
\par Let us consider the multipole expansions of the electrostatic field. Given an electric charge  distribution with support included in the domain $\dom$,   the scalar potential is expressed  in the exterior of a sphere containing  this domain by the following multipolar series:
 \beqa\label{21}
\Phi(\rvec)=\frac{1}{4\pi\eps_0}\suml_{n\ge 0}\frac{(-1)^n}{n!}\d_{i_1}\dots\d_{i_n}\frac{\psf_{i_1\dots i_n}}{r}
=\frac{1}{4\pi\eps_0}\suml_{n\ge 0}\frac{(-1)^n}{n!}\nablav^n\vert\vert\frac{\psft^{(n)}}{r}\ .
\eeqa
In this expansion, the coordinate system origin $O$ is supposed in $\dom$ and
 $\psft^{(n)}$ is  the $n$-th order electric multipolar moment defined by the Cartesian components in the general dynamic case:
\beqa\label{22}
\psf_{i_1\dots i_n}(t)=\int_{\dom}\,\rmd^3x\,\,x_{i_1}\dots x_{i_n}\,\rho(\rvec,t):\;\;\psft^{(n)}(t)=\intl_{\dom}\rmd^3x\,\,\rvec^n\,\rho(\rvec,t).
\eeqa
 In equation \eref{21}  we employed the following notation for tensorial contractions:
\beqa\label{23}
 ({\Atens}^{(n)}||{\Btens}^{(m)})_{i_1 \cdots i_{|n-m|}}
=\left\{\begin{array}{ll}
A_{i_1 \cdots i_{n-m}j_1 \cdots j_m}B_{j_1 \cdots j_m} & ,\; n>m\\
A_{j_1 \cdots j_n}B_{j_1 \cdots j_n} & ,\; n=m\\
A_{j_1 \cdots j_n}B_{j_1 \cdots j_n i_1 \cdots i_{m-n}} & ,\; n<m
\end{array} \right..
\eeqa
For the  multipole  expansion of the electric field $\Evec(\rvec)=-\nablav\,\Phi(\rvec)$, we can write
 \beqa\label{24}
\fl\;\;\;\;\;\Evec(\rvec)=\frac{1}{4\pi\eps_0}\suml_{n\ge 1}\frac{(-1)^{n-1}}{n!}\nablav^{n+1}\vert\vert\frac{\psft^{(n)}}{r}=
\frac{1}{4\pi\eps_0}\evec_i\suml_{n\ge 1}\frac{(-1)^{n-1}}{n!}\d_i\,\d_{i_1}\dots\d_{i_n}\frac{\psf_{i_1\dots i_n}}{r},
\eeqa
where, for simplicity, the electric charged system  is considered neutral ($Q=0$).
 We have to search the singularities of $\Evec$  given by equations \eref{24}. It appears that  cumbersome calculations are involved for higher $n$ if we apply the formulae for higher order  derivatives of  $1/r$ as in the previous section. However, we can employ an invariance property of the electrostatic  field to the substitutions of all moments $\psft^{(n)}$,  for all $n$, by their corresponding symmetric and trace-free {\bf STF} projections $\pct^{(n)}$  \cite{cv-sc,Gonzales,cv02}.
   Retaining the notation $\pvec$  for  the first order moment, this invariance stands  for the invariance of the multipole expansion of the  electrostatic field to the following substitutions:
  \beqa\label{25}
\pvec,\,\psft^{(2)},\,\psft^{(3)},\,\dots\,&\to&\,\pvec,\,\pct^{(2)},\,\pct^{(3)}\dots~;\nonumber\\
\eeqa
 These {\bf STF} tensors can be expressed by the following formula:
  \beqa\label{26}
 \pc_{i_1\dots      i_n}&=&\frac{(-1)^n}{(2n-1)!!}\int_{\dom}\,\rmd^3x\,\rho(\rvec)\rvec^{2n+1}\d_{i_1}\dots \d_{i_n}\frac{1}{r},
  \eeqa
 which, actually, differ from the projections by  numerical factors. Here, 
 \beqan
 \d^{(\la)}_{i_1\dots i_n}=\d_{i_1}\dots \d_{i_{\la-1}}\d_{i_{\la+1}}\dots \d_{i_n}\ .
 \eeqan
 Some care is necessary when one considers this invariance property when we have to establish the  delta-type singularities of the electromagnetic field, since this property is true only for $r\ne 0$. Indeed, as we firstly  see in the electrostatic case, the substitutions $\psft^{(n)}\,\to\,\pct^{(n)}$ in equation \eref{24} give additional terms containing $\Delta(1/r)$ and their derivatives,  which, in the case $r\ne 0$, can be eliminated. But,  extending the expressions associated to the multipole expansions to the entire space, including the origin $O$, these additional terms give notable contributions. This process of searching the delta-form singularities of the electromagnetic field is lost in the previous paper \cite{cvrz-arx} and here, we try to correct it. 
 \par Let us consider the delta-singularity corresponding to the electric dipolar field:
  \beqa\label{27}
  \Evec^{(1)}_{(0)}=\frac{1}{4\pi\eps_0}\evec_i\d_i\d_j\frac{p_j}{r}
=-\frac{1}{3\eps_0}\pvec\,\delta(\rvec),
 \eeqa
a result obtained by directly applying equation \eref{16}.
 \par For the 4-polar term from $\Evec$, we firstly consider the expansion \eref{24} expressed by the primitive moments $\psft^{(n)}$:
  \beqa\label{28}
\Evec^{(2)}(\rvec)=-\frac{1}{8\pi\eps_0}\evec_i\d_i\d_j\d_k\frac{\psf_{jk}}{r}\ .
\eeqa
With the help of equation \eref{20}, we obtain:
 \beqa\label{29}
\fl\Evec^{(2)}_{(0)}&=&\frac{1}{10\eps_0}\evec_i\,\psf_{jk}\delta_{\{ij}\d_{k\}}\delta(\rvec)
=\frac{1}{10\eps_0}\evec_i\,\psf_{jk}\left(\delta_{ij}\d_k\delta(\rvec)+\delta_{ik}\d_j\delta(\rvec)+\delta_{jk}\d_i\delta(\rvec)\right)\nonumber\\
\fl &=&\frac{1}{10\eps_0}\evec_i\left(2\psf_{ij}\d_j\delta(\rvec)+\psf_{jj}\d_i\delta(\rvec)\right)=\frac{1}{5\eps_0}\evec_i\left(\psf_{ij}\d_j\delta(\rvec)+\frac{1}{2}\psf_{jj}\d_i\delta(\rvec)\right).
\eeqa

Note that $\psft^{(2)}$ is symmetric. This result suggests the possibility to facilitate the calculation of the delta-type singularities of the electromagnetic field if instead of the ``primitive'' tensors $\psft^{(n)}$ we can employ the {\bf STF} projections $\pct^{(n)}$. For the present case, it will be simpler to calculate the contractions of the last type of tensors with the coefficients $C^{(n)}$ from equation \eref{6}.
\par For $n=2$, we search the trace-free part of the symmetric tensor $\psft^{(2)}$ as
\beqa\label{30}
\psf_{ij}=\pc_{ij}+\delta_{ij}\,\Lambda \ .
\eeqa
The parameter $\Lambda$ is determined such that $\pc_{ii}=0$. One obtains
\beqa\label{31}
\Lambda=\frac{1}{3}\psf_{ii}\ .
\eeqa
Let us introduce equation \eref{30} in equation \eref{28}:
\beqan
\Evec^{(2)}(\rvec)=-\frac{1}{8\pi\eps_0}\evec_i\,\pc_{jk}\d_i\d_j\d_k\frac{1}{r}-\frac{\Lambda}{8\pi\eps_0}\evec_i\,\d_i\left(\Delta\frac{1}{r}\right)\ .
\eeqan
For $r\ne 0$, the last term vanishes and, indeed, $\Evec$ is invariant to the substitution $\pct^{(2)}\,\to\,\pc^{(2)}$.  Extending this field to the entire space, this last term contributes with a delta-singularity  such that we must write
\beqa\label{32}
\Evec^{(2)}(\rvec)=-\frac{1}{8\pi\eps_0}\evec_i\,\pc_{jk}\d_i\d_j\d_k\frac{1}{r}
+\frac{1}{2\eps_0}\Lambda\,\nablav\delta(\rvec)\ .
\eeqa
Inserting equation \eref{20} in equation \eref{32}, and retaining only the delta-singularities, we can write
\beqa\label{33}
\Evec^{(2)}_{(0)}(\rvec)=\frac{1}{5\eps_0}\,\evec_i\pc_{ij}\d_j\delta(\rvec)
+\frac{1}{2\eps_0}\Lambda\,\nablav\delta(\rvec)\ .
\eeqa
Substituting equation \eref{30} in equation \eref{33}, one obtains equation \eref{29}. 
\par The advantage  of employing the {\bf STF} moments  $\pct^{(n)}$ instead of the primitive moments $\psft^{(n)}$ is
manifest for higher order terms from the multipolar expansion. Even from $n=3$, the contraction of the angular average of a product of more then six factors $\nu$ with a primitive moment represented by a tensor which is only symmetric becomes cumbersome.
\par Though, maybe, only of theoretical interest, let us consider the general case of arbitrary $n$. The {\bf STF} projection $\Tcal(\pct^{(n)})$ of the symmetric tensor $\psft^{(n)}$ is defined, up to a numerical factor, by the equation
\beqa\label{34}
\Tcal_{i_1\dots i_n}(\psft^{(n)})\equiv \pc_{i_1\dots i_n}=\psf_{i_1\dots i_n} -\delta_{\{i_1i_2}\lasf_{i_3\dots i_n\}} \ ,
\eeqa
where $\latens^{(n-2)}$ is a symmetric tensor and is defined by the conditions of the trace-free character of $\pct^{(n)}$. For low values of $n$ (the ones of practical interest),  the components $\lasf_{i_1\dots i_n}$ can be calculated directly from the equation system representing the vanishing of all the traces of the tensor $\pct^{(n)}$. For higher orders $n$, there is a general formula known in literature \cite{Thorne}, \cite{ap} which, with the notation from the present paper, is written as
\beqa\label{35}
&~&\fl\left[\Tcal\big[\psft^{(n)}\big]\right]_{i_1\dots i_n}=
\suml^{[n/2]}_{m=0}\frac{(-1)^m(2n-1-2m)!!}{(2n-1)!!}\delta_{\{i_1i_2}\dots
\delta_{i_{2m-1}i_{2m}}\psf^{(n:m)}_{i_{2m+1}\dots i_n\}} \ .
\eeqa
 $\psf^{(n:m)}_{i_{2m+1}\dots i_n}$ denotes the components of the $(n-2\,m)$-th order tensor 
obtained from $\psft^{(n)}$ by contracting $m$ pairs of symbols $i$. This equation is known as the {\it detracer theorem} \cite{ap}. As a consequence of this theorem, the components of the tensor $\latens^{(n-2)}$ are written as
\beqa\label{36}
\fl\lasf_{i_1\dots i_{n-2}}\big[\psft^{(n)}\big]=
\suml^{[n/2-1]}_{m=0}\frac{(-1)^m[2n-1-2(m+1)]!!}{(m+1)(2n-1)!!}
\delta_{\{i_1i_2}\dots \delta_{i_{2m-1}i_{2m}}\psf^{(n:\,m+1)}_{i_{2m+1}\dots i_{n-2}\}}.
\eeqa
Inserting equation \eref{34} in the expression of $\Evec^{(n)}(\rvec)$ given by equation \eref{24}, we obtain
\beqan
\Evec^{(n)}(\rvec)=\frac{(-1)^{n-1}}{4\pi\eps_0n!}\pct^{(n)}\vert\vert\nablav^{n+1}\frac{1}{r}
+\frac{(-1)^{n-1}}{4\pi\eps_0n!}\evec_i\delta_{\{i_1i_2}\lasf_{i_3\dots i_n\}}\,\d_i\d_{i_1}\dots\d_{i_n}\frac{1}{r} \ .
\eeqan
and, since the contraction with the fully symmetric tensor $\latens^{(n-2)}$ produces $C^2_n=n(n-1)/2$ identical terms,
\beqa\label{37}
\fl \Evec^{(n)}(\rvec)&=&\frac{(-1)^{n-1}}{4\pi\eps_0n!}\pct^{(n)}\vert\vert\nablav^{n+1}\frac{1}{r}
+\frac{(-1)^{n-1}}{8\pi\eps_0(n-2)!}\evec_i\lasf_{i_1\dots i_{n-2}}\d_i\d_{i_1}\dots \d_{i_{n-2}}\Delta\frac{1}{r}\nonumber\\
\fl&=&\frac{(-1)^{n-1}}{4\pi\eps_0n!}\pct^{(n)}\vert\vert\nablav^{n+1}\frac{1}{r}+\frac{(-1)^n}{2(n-2)!\,\eps_0}\latens^{n-2}\vert\vert\nablav^{n-1}\delta(\rvec) \ .
\eeqa
Searching the extension of $\Evec^{(n)}$ to the entire space, for establishing the $\delta$-type singularities, we have to calculate the limit
\beqan
\left\langle\Evec^{(n)}_{(0)},\,\phi\right\rangle&=&\frac{(-1)^{n-1}}{4\pi\eps_0\,n!}\lim_{\eps\to 0}\
\int_{\dom_\eps}\rmd^3x\,\left(\pct^{(n)}\vert\vert\nablav^{n+1}\frac{1}{r}\right)\,\phi(\rvec)\\
&+&\frac{(-1)^n}{2(n-2)!\,\eps_0}\lim_{\eps\to 0}\int_{\dom_\eps}\rmd^3x\,\latens^{(n-2)}\vert\vert\nablav^{n-1}\delta(\rvec)\,\phi(\rvec) \ ,
\eeqan 
or
\beqa\label{38}
\left\langle\Evec^{(n)}_{(0)},\,\phi\right\rangle&=&\frac{(-1)^{n-1}}{4\pi\eps_0\,n!}\lim_{\eps\to 0}\
\int_{\dom_\eps}\rmd^3x\,\left(\pct^{(n)}\vert\vert\nablav^{n+1}\frac{1}{r}\right)\,\phi(\rvec)\nonumber\\
&-&\frac{1}{2(n-2)!\eps_0}\latens^{((n-2)}\vert\vert\left(\nablav^{n-1}\phi\right)_0 \ .
\eeqa
As done in Ref. \cite{cvrz-arx},  the limit of the remaining integral from equation \eref{38} can be easily expressed for arbitrary $n$. Applying the Gauss theorem, we can write
\beqa\label{38a}
\fl\lim_{\eps\to 0}\int_{\dom_\eps}\rmd^3x\,\left(\pct^{(n)}\vert\vert\nablav^{n+1}\frac{1}{r}\right)\,\phi(\rvec)\nonumber\\
\fl =\lim_{\eps\to 0}\evec_i\left[\oint_{\Sigma_{\eps}}\rmd S\,\nu_i\,\left(\pc_{i_1\dots i_n}\d_{i_1}\dots \d_{i_n}\frac{1}{r}\right)\,\phi(\rvec)-\int_{\dom_\eps}\rmd^3x\,\left(\pc_{i_1\dots i_n}\d_{i_1}\dots \d_{i_n}\frac{1}{r}\right)\,\d_i\phi(\rvec)\right].
\eeqa
Introducing the Taylor series for the function $\phi(\rvec)$ in the surface integral from the last equation and since
$r=\eps$ on the sphere $\Sigma_\eps$,  we can write
\beqa\label{39}
\fl\lim_{\eps\to 0}\evec_i\oint_{\Sigma_{\eps}}\rmd S\,\nu_i\,\left(\pc_{i_1\dots i_n}\d_{i_1}\dots \d_{i_n}\frac{1}{r}\right)\,\phi(\rvec)\nonumber\\
\fl\;\;\;\;\;\;\;\;\;\;\;\;\;=4\pi\,\evec_i\lim_{\eps\to 0}\suml^\infty_{\al
=0}\frac{\eps^{\al-n+1}}{\al!}\pc_{i_1\dots i_n}\left\langle\,\nu_i\,C^{(n)}_{i_1\dots i_n}\nu_{i_{n+1}}\dots \nu_{i_{n+\al}}\right\rangle\,\left(\d_{i_{n+1}}\dots \d_{i_{n+\al}}\,\phi\right)_0\ .
\eeqa
Let us evaluate the tensorial contraction which is present of the general term in the series from the previous equation:
\beqan
\pc_{i_1\dots i_n}\left\langle\,\nu_i\,C^{(n)}_{i_1\dots i_n}\,\nu_{i_{n+1}}\dots \nu_{i_{n+\al}}\right\rangle\ .
\eeqan
From equation \eref{7} we can easily see that all the terms containing at least a symbol $\delta_{i_qi_s}$ with $1\le q,\,s\le n$ give null results by contraction with the traceless tensor $\pc^{(n)}$. Only the term corresponding to $k=0$ in equation \eref{7} can give results different from zero for this contraction. Moreover, a result different from zero can be obtained if and only if $\al+1\ge n$. Therefore, equation \eref{41} can be written as
\beqa\label{40}
\fl\lim_{\eps\to 0}\evec_i\oint_{\Sigma_{\eps}}\rmd S\,\nu_i\,\left(\pc_{i_1\dots i_n}\d_{i_1}\dots \d_{i_n}\frac{1}{r}\right)\,\phi(\rvec)
 = 4\pi\,\evec_i(-1)^n(2n-1)!!\nonumber\\
\fl \;\;\;\;\;\;\;\;\;\;\;\;\;\;\times\lim_{\eps\to 0}\suml^\infty_{\al=n-1}\frac{\eps^{\al-n+1}}{\al!}\pc_{i_1\dots i_n}\left\langle\,\nu_{i_1}\dots \nu_{i_n}\nu_{i_{n+1}}\dots \nu_{i_{n+\al}}\,\nu_i\right\rangle\,\left(\d_{i_{n+1}}\dots \d_{i_{n+\al}}\,
\phi\right)_0\ .
\eeqa
  For $\al\,>\,n-1$, the corresponding terms from the series in equation \eref{40} contain positive powers of $\eps$. Consequently, the corresponding limits for $\eps\,\to 0$  vanish. 
  From this series only the term for which 
  \beqa\label{41}
 \al=n-1
 \eeqa
 can be different from zero and the result of the limit in equation \eref{40} is given by
  \beqa\label{42}
&&\fl \lim_{\eps\to 0}\evec_i\oint_{\Sigma_{\eps}}\rmd S\,\nu_i\,\left(\pc_{i_1\dots i_n}\d_{i_1}\dots \d_{i_n}\frac{1}{r}\right)\,\phi(\rvec)
=\frac{4\pi(-1)^n(2n-1)!!}{(n-1)!}  \nonumber\\
&& \times\evec_i\,\pc_{i_1\dots i_n}\left\langle\,\nu_{i_1}\dots \nu_{i_n}\,\nu_{i_{n+1}}\dots \nu_{i_{2n-1}}\,\nu_i\right\rangle\,\left(\d_{i_{n+1}}\dots \d_{i_{2n-1}}\,\phi\right)_0\nonumber\\
\fl&=&\frac{4\pi(-1)^n(2n-1)!!}{(n-1)!}\left\langle\,\pct^{(n)}\vert\vert\nuvec^{2n}\right\rangle\vert\vert\left(\nablav^{n-1}\phi\right)_0\ .
 \eeqa

 \par Let us evaluate now the volume integral in equation \eref{38a}:
  \beqa\label{44}
\fl\;\;\;\;\;\;&&\lim_{\eps\to0}\int_{\dom_\eps}\rmd^3x\,\left(\nablav^n\vert\vert\frac{\pct^{(n)}}{r}\right)\,\nablav\phi(\rvec)\nonumber\\
 \fl&&\ \ \ =\lim_{\eps\to 0}\left[\oint_{\Sigma_\eps}\rmd S\,\nuvec\vert\vert\left(\nablav^{n-1}\vert\vert\frac{\pct^{(n)}}{r}\right)\,\nablav\phi(\rvec)
 -\int_{\dom\eps}\rmd^3x\,\left(\nablav^{n-1}\vert\vert\frac{\pct^{(n)}}{r}\right)\,\nablav^2\phi(\rvec) \right]\ .
 \eeqa
 The part corresponding to the surface integral can be written as
  \beqa\label{45}
\fl\lim_{\eps\to 0}\oint_{\Sigma_\eps}\rmd S\,\nuvec\vert\vert\left(\nablav^{n-1}\vert\vert\frac{\pct^{(n)}}{r}\right)\,\nablav\phi(\rvec)
=\evec_i\pc_{i_1\dots i_n}\lim_{\eps\to 0}\oint_{\Sigma_\eps}\rmd S\,\nu_{i_n}\,\left(\d_{i_1}\dots \d_{i_{n-1}}\frac{1}{r}\right)\,\d_i\phi(\rvec)\ .\nonumber\\
 \eeqa
 Introducing the Taylor series for $\phi(\rvec)$ and standing out the average over $\nuvec$, we obtain
  \beqa\label{46}
 \fl\lim_{\eps\to 0}\oint_{\Sigma_\eps}\rmd S\,\nuvec\vert\vert\left(\nablav^{n-1}\vert\vert\frac{\pct^{(n)}}{r}\right)\,\nablav\phi(\rvec)\nonumber\\
\fl\;\;\;\;\;\;=4\pi\,\evec_i\lim_{\eps\to 0}\suml^\infty_{\al=0}\frac{\eps^{\al-n+2}}{\al!}\pc_{i_1\dots i_n}\left\langle\,C^{(n-1,n-1)}_{i_1\dots i_{n-1}}\nu_{i_n}\dots \nu_{i_{n+\al}} \right\rangle
 \times \left(\d_{i_{n+1}}\dots \d_{i_{n+\al}}\d_i\phi\right)_0.
 \eeqa
Analogously to the reasoning from the previous case, we can see that the limit is zero since for $\al=n-2$ 
 \beqan
 \pc_{i_1\dots i_n}\left\langle\,C^{(n-1,n-1)}_{i_1\dots i_{n-1}}\nu_{i_n}\nu_{i_{n+1}}\dots \nu_{i_{2n-2}}
 \right\rangle=0.
 \eeqan
 Therefore, 
  \beqa\label{47}
\fl\lim_{\eps\to0}\int_{\dom_\eps}\rmd^3x\,\left(\nablav^n\vert\vert\frac{\pct^{(n)}}{r}\right)\,\nablav\phi(\rvec)=-\lim_{\eps\to 0}\int_{\dom_\eps}\rmd^3x\,\left(\nablav^{n-1}\vert\vert\frac{\pct^{(n)}}{r}\right)\,\nablav^2\phi(\rvec)\ ,
\eeqa 
and by repeatedly applying the procedure, all terms cancel. Finally, only the surface integral from equation \eref{38a} gives a limit different from zero and
 \beqa\label{48}
 \fl\;\;\lim_{\eps\to 0}\int_{\dom_\eps}\rmd^3x\,\left(\pct^{(n)}\vert\vert\nablav^{n+1}\frac{1}{r}\right)\,\phi(\rvec)=
\frac{4\pi(-1)^n(2n-1)!!}{(n-1)!}\left\langle\,\pct^{(n)}\vert\vert\nuvec^{2n}\right\rangle\vert\vert\left(\nablav^{n-1}\phi\right)_0\ .
 \eeqa
 Let us consider the contraction 
 \beqan
 \left\langle\,\pct^{(n)}\vert\vert\nuvec^{2n}\right\rangle=\pc_{i_1\dots i_n}\,\left\langle\, \nu_{i_1}\dots\nu_{i_n}\nu_{j_1}\dots\nu_{j_n}\right\rangle\ .
 \eeqan
 To this contraction contribute only the terms from the average of the $\nu$-product not containing factors $\delta_{i_ki_l}$, with $1\le k,\,l\le n$. According to equation \eref{15}, the terms giving non-zero contributions are of the form
 \beqan
 \frac{1}{(2n+1)!!}\,\delta_{i_1j_1}\dots \delta_{i_nj_n}
   \eeqan
  and all such terms are obtained considering  the $n!$ permutations of the indices $j_1\dots j_n$ in this product. Therefore, the final expression in equation \eref{42} is given by
\beqa\label{49}
\fl\;\;\;\;\;\;\lim_{\eps\to 0}\int_{\dom_\eps}\rmd^3x\,\left(\pct^{(n)}\vert\vert\nablav^{n+1}\frac{1}{r}\right)\,\phi(\rvec)   =\frac{4\pi(-1)^n\,n}{2n+1}\,\pct^{(n)}\vert\vert\left(\nablav^{n-1}\phi\right)_0\ .
\eeqa
Equation \eref{38}, with equation \eref{49} inserted in, yields
 \beqa\label{50}
\fl\left\langle\,\Evec^{(n)}_{(0)},\,\phi\right\rangle=-\frac{1}{(n-1)!(2n+1)\eps_0}\pct^{(n)}\vert\vert\left(\nablav^{n-1}\phi\right)_0-\frac{1}{2(n-2)!\eps_0}\latens^{(n-2)}\vert\vert\left(\nablav^{n-1}\phi\right)_0 .
\eeqa
Therefore, the delta-form distribution associated to the $2^n$-polar electric field is given by
 \beqa\label{51}
 \fl\;\;\;\;\;\;\Evec^{(n)}_{(0)}
 =\frac{(-1)^n}{(n-1)!\,(2n+1)\,\eps_0}\pct^{(n)}\vert\vert\nablav^{n-1}\delta(\rvec)
 +\frac{(-1)^n}{2(n-2)!\,\eps_0}\latens^{(n-2}\vert\vert\nablav^{n-1}\delta(\rvec)\ .
 \eeqa
One can easily see that the result \eref{32} for $n=2$ is, indeed,  a particular case of the formula \eref{51}. In the case $n=3$, equation \eref {51} becomes:
\beqan
\Evec^{(3)}_{(0)}=-\frac{1}{14\eps_0}\pct^{(3)}\vert\vert\nablav^2\delta(\rvec)-\frac{1}{2\eps_0}\latens^{(1)}\vert\vert\nablav^2\delta(\rvec)\ .
\eeqan
The result can be expressed in terms of primitive tensors $\psft^{(n)}$ introducing equation \eref{34} in equation \eref{51}:
\beqan
\fl\Evec^{(n)}_{(0)}
 &=&\frac{(-1)^n}{(n-1)!\,(2n+1)\,\eps_0}\psft^{(n)}\vert\vert\nablav^{n-1}\delta(\rvec)
 +\frac{(-1)^n}{2(n-2)!\,\eps_0}\latens^{(n-2}\vert\vert\nablav^{n-1}\delta(\rvec)\\
 \fl &-&\frac{(-1)^n}{(n-1)!(2n+1)\eps_0}\evec_i\delta_{\{i\,i_1}\lasf_{i_2\dots i_{n-1}\}}\d_{i_1}\dots \d_{i_{n-1}}\delta(\rvec) \ .
\eeqan
In the last  term from this equation, there are $n-1$ identical terms of the form
\beqan
\evec_i\lasf_{i_2\dots i_{n-1}}\d_i\,\d_{i_2}\dots \d_{i_{n-1}}\delta(\rvec)=\latens^{(n-2)}\vert\vert\nablav^{n-1}\delta(\rvec)
\eeqan
and $C^2_{n-1}=(n-1)(n-2)/2$ terms of the form
\beqan
\evec_i\lasf_{i\,i_3\dots i_{n-1}}\d_{i_3}\dots \d_{i_{n-1}}\delta(\rvec)=\latens^{(n-2)}\vert\vert\nablav^{n-3}\Delta\delta(\rvec) \ .
\eeqan
Finally,
    \beqa\label{52} 
 \fl \Evec^{(n)}_{(0)}=\frac{(-1)^n}{(n-1)!\,(2n+1)\,\eps_0}\psft^{(n)}\vert\vert\nablav^{n-1}\delta(\rvec)
 +\frac{(-1)^n(2n-1)}{2(n-2)!(2n+1)\eps_0}\latens^{(n-2)}\vert\vert\nablav^{n-1}\delta(\rvec)\nonumber\\
 -\frac{(-1)^n}{2(n-3)!(2n+1)\eps_0}\latens^{(n-2)}\vert\vert\nablav^{n-3}\,\Delta\delta(\rvec)\ .
 \eeqa
This last equation becomes equation \eref{29} in case $n=2$. The terms containing $\Delta \delta(\rvec)$ are present beginning from $n=3$.

\section{Singularities of the magnetostatic field}
 For the vector potential in the exterior of the domain $\dom$, we have
   \beqa\label{53}
\Avec(\rvec)&=&\frac{\mu_0}{4\pi}\suml_{n\ge 1}\frac{(-1)^{n-1}}{n!}\nablav\times\left(\nablav^{n-1}\vert\vert\msft^{(n)}\right)\nonumber\\
&=&\frac{\mu_0}{4\pi}\evec_i\eps_{ijk}\d_j\suml_{n\ge 1}\frac{(-1)^{n-1}}{n!}\d_{i_1}\dots\d_{i_{n-1}}\frac{\msf_{i_1\dots i_{n-1}\,k}}{r},
\eeqa
where $\msft^{(n)}$ is the magnetic $n$-th order moment defined by the Cartesian components  \cite{Castell}:
 \beqa\label{54}
 \msf_{i_1\dots i_n}(t)=\frac{n}{n+1}\int_{\dom}\rmd^3x\,\,x_{i_1}\dots x_{i_{n-1}}\big(\rvec\times\jvec(\rvec,t)\big)_{i_n},
 \eeqa
 or, with tensorial notation:
 \beqan
 \msft^{(n)}(t)=\frac{n}{n+1}\intl_{\dom}\rmd^3x\,\,\rvec^n\times\jvec(\rvec,t)\ .
\eeqan

\par The corresponding expansion of the magnetic field $\Bvec(\rvec)=\nablav\times\Avec(\rvec)$ is given by
  \beqa\label{55}
\fl \;\;\;\;\;\;\;\;\;\;\;\;\;\;\;\;\;\;\;\;\;\;\Bvec(\rvec)&=&\frac{\mu_0}{4\pi}\nablav\times\suml_{n\ge 1}\frac{(-1)^{n-1}}{n!}\nablav\times\left(\nablav^{n-1}\vert\vert\frac{\msft^{(n)}}{r}\right)\nonumber\\
\fl&=&\frac{\mu_0}{4\pi}\suml_{n\ge 1}\frac{(-1)^{n-1}}{n!}\left[\nablav\cdot\left(\nablav^n\vert\vert\frac{\msft^{n)}}{r}\right)
-\Delta\left(\nablav^{n-1}\vert\vert\frac{\msft^{(n)}}{r}\right)\right]\ .
 \eeqa
 For $r\ne 0$, the last term containing $\Delta(1/r)$ is not contributing to the multipole expansion since $\Delta(1/r)=0$, but, searching the extension of this expansion to the entire space, including the point $O$, we have to consider it. This term, extended as in the electrostatic case,  as $4\pi\,\msft^{(n)}\vert\vert\nablav^{n-1}\delta(\rvec)$, is considered as a first extension of $\Bvec$. It remains to process the limit that implies the expression $\nablav\left(\msft^{(n)}\vert\vert\nablav^n(1/r)\right)$.\\
 In the dipolar case, we write
 \beqan
 \Bvec^{(1)}=\frac{\mu_0}{4\pi}\evec_i\d_i\d_j\frac{m_j}{r}+4\pi\,\mvec\delta(\rvec)
 \eeqan
and, applying equation \eref{16}, we obtain for the delta-singularity, the well-known expression \cite{Jackson}:
  \beqa\label{56}
\Bvec^{(1)}_{(0)}=-\frac{\mu_0}{3}\mvec\,\delta(\rvec)+\mu_0\mvec\,\delta(\rvec)=\frac{2\mu_0}{3}\mvec\,\delta(\rvec)\ .
\eeqa
For the higher multipolar orders, let us begin with the 4-polar term:
\beqan
\Bvec^{(2)}=-\frac{\mu_0}{8\pi}\nablav\left(\nablav^2\vert\vert\frac{\msft^{(2)}}{r}\right)=-\frac{\mu_0}{8\pi}\evec_i\msf_{jk}\d_i\d_j\d_k\frac{1}{r}-\frac{\mu_0}{2}\evec_i\msf_{ji}\d_j\delta(\rvec) \ .
\eeqan
Applying equation \eref{20},
\beqan
\fl \Bvec^{(2)}_{(0)}=\frac{\mu_0}{10}\evec_i\msf_{jk}\delta_{\{ij}\d_{k\}}\,\delta(\rvec)-\frac{\mu_0}{2}\evec_i\msf_{ji}\d_j\delta(\rvec)=\frac{\mu_0}{10}\evec_i\left(\msf_{ij}+\msf_{ji}\right)\d_j\delta(\rvec)-\frac{\mu_0}{2}\evec_i\msf_{ji}\d_j\delta(\rvec)
\eeqan
since $\delta_{jk}\msf_{jk}=\msf_{jj}=0$. Further,
  \beqa\label{57}
\fl\Bvec^{(2)}_{(0)}=\frac{\mu_0}{5}\evec_i\mlr_{ij}\d_j\delta(\rvec)-\frac{\mu_0}{2}\evec_i\msf_{ji}\d_j\delta(\rvec)
=\frac{\mu_0}{5}\mlrt^{(2)}\vert\vert\nablav\delta(\rvec)-\frac{\mu_0}{2}\left(\nablav\delta(\rvec)\right)\vert\vert\msft^{(2)}\ ,
\eeqa
where $\mlrt^{(2)}$ is the symmetric part of the tensor $\msft^{(2)}$ corresponding to the identity
  \beqa\label{58}
\msf_{ij}=\frac{1}{2}\left(\msf_{ij}+\msf_{j\,i}\right)+\frac{1}{2}\left(\msf_{ij}-\msf_{j\,i}\right)=\mlr_{ij}+\frac{1}{2}\eps_{ijk}\Nsf_k \ .
\eeqa
The antisymmetric part is expressed in terms of the components of a first rank tensor:
\beqan
\Nsf_i=\eps_{ijk}\msf_{jk}=\frac{2}{3}\int_{\dom}\rmd^3x\,\left[\rvec\times(\rvec\times\jvec)\right]_i\ .
\eeqan
For $n=2$, the symmetric projection of the magnetic moment is an {\bf STF} tensor, i.e, denoting by $\mct^{(n)}$ the {\bf STF} magnetic moments, 
\beqan
\mct^{(2)}=\mlrt^{(2)} \ ,
\eeqan
the equation \eref{58} is written as
 \beqa\label{59}
 \msf_{ij}=\mc_{ij}+\frac{1}{2}\eps_{ijk}\Nsf_k\ .
 \eeqa
The introduction of   these result in equation \eref{57} gives
  \beqa\label{60}
 \Bvec^{(2)}_{(0)}=-\frac{3\mu_0}{10}\mct^{(2)}\vert\vert\nablav\delta(\rvec)-\frac{\mu_0}{4}\bbox{N}\times\nablav\delta(\rvec) \ ,
 \eeqa
 where $\bbox{N}=\Nsf_i\evec_i$.\\
 \par Beginning from $n=3$, the symmetric part of the magnetic moment tensor is not the same with the {\bf STF} one. We write the identity
  \beqa\label{61} \fl\msf_{i_1i_2i_3}=\frac{1}{3}\left(\msf_{i_1i_2i_3}+\msf_{i_2i_3i_1}+\msf_{i_1i_3i_2}\right)+\frac{1}{3}\left[\left(\msf_{i_1i_2i_3}-\msf_{i_2i_3i_1}\right)+
 \left(\msf_{i_1i_2i_3}-\msf_{i_1i_3i_2}\right)\right] \ ,
 \eeqa
 where $\msf_{i_1i_2i_3}$ is symmetric in the first two indices. The first parenthesis represents the fully symmetric part of the tensor $\msft^{(3)}$ and the second one can be expressed in terms of the second order tensor 
 $\Nsft^{(2)}$ defined by the components
 \beqan
 \Nsf_{i_1i_2}=\eps_{i_2pq}\msf_{i_1pq}=\frac{3}{4}\int_{\dom}\rmd^3x\,x_{i_1}\left[\rvec\times(\rvec\times\jvec)\right]_{i_2} \ ,
 \eeqan
 with the relationships:
 \beqan
 \msf_{i_1i_2i_3}-\msf_{i_2i_3i_1}=\eps_{i_1i_3q}\Nsf_{i_2q},\;\;
 \msf_{i_1i_2i_3}-\msf_{i_1i_3i_2}=\eps_{i_2i_3q}\Nsf_{i_1q}\ .
 \eeqan
We can write
\beqan
\fl\;\;\;\;\;\;\;\;\left(\msf_{i_1i_2i_3}-\msf_{i_2i_3i_1}\right)+ \left(\msf_{i_1i_2i_3}-\msf_{i_1i_3i_2}\right)=\eps_{i_1i_3q}\Nsf_{i_2q}+\eps_{i_2i_3q}\Nsf_{i_1q}=
\suml^2_{\la=1}\eps_{i_\la i_3q}\,\Nsf^{(\la)}_{(i_1i_2q)} \ ,
\eeqan
where by the notation $\Nsf^{(\la)}_{(i_1i_2q)}$ we understand the component without the index $i_\la$. Employing the above definitions and notation, equation \eref{61} can be written as
 \beqa\label{62}
\msf_{i_1i_2i_3}=\mlr_{i_1i_2i_3}+\frac{1}{3}\suml^2_{\la=1}\eps_{i_\la i_3q}\Nsf^{(\la)}_{(i_1i_2q)} \ ,
\eeqa
where $\mlrt^{(3)}$ is the symmetric part of the tensor $\msft^{(3)}$.\\
Let us write the  third order term from the magnetic field expansion \eref{55} 

 \beqan
\Bvec^{(3)}(\rvec)=\frac{\mu_0}{24\pi}\nablav\left(\nablav^3\vert\vert\frac{\msft^{(3)}}{r}\right)-\frac{\mu_0}{24\pi}\nablav^2\vert\vert\left(\Delta\frac{\msft^{(3)}}{r}\right) \ .
\eeqan
Retaining the second expression which represent an extension as distribution with the point-like support $O$:
\beqa\label{63}
\Bvec^{(3)}(\rvec)=\frac{\mu_0}{24\pi}\nablav\left(\nablav^3\vert\vert\frac{\msft^{(3)}}{r}\right)
+\frac{\mu_0}{6}\left(\nablav^2\delta(\rvec)\right)\vert\vert\msft^{(3)}\ .
\eeqa
Introducing the symmetric tensor $\mlrt^{(3)}$ from equation \eref{62}  in the first expression of the right-hand side of the above equation,
\beqan
\nablav\left(\nablav^3\vert\vert\frac{\msft^{(3)}}{r}\right)=\nablav\left(\nablav^3\vert\vert\frac{\mlrt^{(3)}}{r}\right)+\frac{1}{3}\evec_i\d_i\d_{i_1}\d_{i_2}\d_{i_3}\frac{1}{r}\suml^2_{\la=1}\eps_{i_\la  i_3q}\Nsf^{(\la)}_{(i_1i_2q)}\ .
\eeqan
Since in the last expression, the two terms of the sum contain either the  $\d_{i_1}\d_{i_3}\eps_{i_1i_3q}$ 
or $\d_{i_2}\d_{i_3}\eps_{i_2i_3q}$ which vanish, we can write 
\beqa\label{64}
\nablav\left(\nablav^3\vert\vert\frac{\msft^{(3)}}{r}\right)=\nablav\left(\nablav^3\vert\vert\frac{\mlrt^{(3)}}{r}\right)\ .
\eeqa
This result expresses the invariance of the multipole expansion of the magnetic field to the substitution $\msft^{(3)}\,\to \,\mlrt^{(3)}$. The introduction of $\mlrt^{(3)}$ in the $\delta$-type singularity from equation \eref{63} gives
\beqan
\fl&&\left(\nablav^2\delta(\rvec)\right)\vert\vert\msft^{(3)}
=\left(\nablav^2\delta(\rvec)\right)\vert\vert\mlrt^{(3)}+\frac{1}{3}\evec_i\d_{i_1}\d_{i_2}\delta(\rvec)\suml^2_{\la=1}\eps_{i_\la iq}\Nsf^{(\la)}_{(i_1i_2q)}\\
\fl&=&\left(\nablav^2\delta(\rvec)\right)\vert\vert\mlrt^{(3)}+\frac{2}{3}\evec_i\Nsf_{i_2q}\eps_{iqi_1}\d_{i_1}\d_{i_2}\delta(\rvec)=\mlrt^{(3)}\vert\vert\nablav^2\delta(\rvec)+\frac{2}{3}\evec_i\Nsf_{i_2q}\eps_{iqi_1}\d_{i_1}\d_{i_2}\delta(\rvec)
\eeqan
since $\mlrt$ is  symmetric.
For expressing in a compact form such tensorial contraction, let us introduce the notation 
\beqa\label{65}
\Asft^{(n)}\vert\times\vert\Bsft^{(n)}=\evec_i\Asf_{i_1\dots i_{n-1}q}\eps_{iqs}\Bsf_{s\,i_1\dots i_{n-1}} \ .
\eeqa
Then, 
\beqa\label{66}
\left(\nablav^2\delta(\rvec)\right)\vert\vert\msft^{(3)}=\,\mlrt^{(3)}\vert\vert\nablav^2\delta(\rvec)
+\frac{2}{3}\Nsft^{(2)}\vert\times\vert\nablav^2\delta(\rvec)\ .
\eeqa
The octupolar term $\Bvec^{(3)}$ including partially $\delta$-form singularities becomes
\beqa\label{67}
\fl\;\;\;\;\;\;\Bvec^{(3)}(\rvec)=\frac{\mu_0}{24\pi}\nablav\left(\nablav^3\vert\vert\frac{\mlrt^{(3)}}{r}\right)+\frac{\mu_0}{6}\mlrt^{(3)}\vert\vert\nablav^2\delta(\rvec)
+\frac{\mu_0}{9}\Nsft^{(2)}\vert\times\vert\nablav^2\delta(\rvec)\ .
\eeqa
The {\bf STF} projection $\mct^{(3)}$, up to a numerical factor, is given by
\beqa\label{68}
\mc_{i_1i_2i_3}=\mlr_{i_1i_2i_3}-\delta_{\{i_1i_2}\widetilde{\lasf}_{i_3\}}(\mlrt^{(3)}) \ ,
\eeqa
where the symmetric tensor $\widetilde{\latens}^{(n-2)}$ corresponds to $\msft^{(n)}$ by the a formula of the type \eref{34}.
It easy to see that 
\beqa\label{69}
\widetilde{\lasf}_i=\frac{1}{5}\mlr_{iqq}=\frac{1}{15}\msf_{qqi}=\frac{1}{20}\int_{\dom}\rmd^3x\,r^2\,(\rvec\times\jvec)_i\ .
\eeqa
The introduction of equation \eref{68} in equation \eref{67} gives
\beqa\label{70}
\fl\Bvec^{(3)}(\rvec)&=&\frac{\mu_0}{24\pi}\nablav\left(\nablav^3\vert\vert\frac{\mct^{(3)}}{r}\right)+\frac{\mu_0}{6}\mct^{(3)}\vert\vert\nablav^2\delta(\rvec)-\frac{\mu_0}{6}\widetilde{\latens}\vert\vert\nablav^2\delta(\rvec)+\frac{\mu_0}{6}\widetilde{\latens}\,\Delta\delta(\rvec)\nonumber\\
\fl&+&\frac{\mu_0}{9}\Nsft^{(2)}\vert\times\vert\nablav^2\delta(\rvec)\ .
\eeqa
\par From this last equation, it is seen that the multipole expansion of $\Bvec(\rvec)$  ($r\ne 0$) is invariant to the substitution $\msft^{(3)}\,\to\,\mct^{(3)}$ \cite{cv-sc,Gonzales}.
\par It remains to calculate the extension of the first term from equation \eref{70}  to the entire space. For this, we have the result \eref{49} from the case of electrostatic field which,  for arbitrary $n$, gives in the present case:
\beqa\label{71}
\lim_{\eps\to 0}\int_{\dom_\eps}\rmd^3x\,\nablav\left(\nabla^n\vert\vert\frac{\mct^{(n)}}{r}\right)\,\phi(\rvec)=\frac{4\pi(-1)^n\,n}{2n+1}\mct^{(n)}\vert\vert\left(\nablav^{n-1}\phi\right)_0 \ .
\eeqa
The final result for the singular part of $\Bvec^{(3)}$ having as support the point $O$ is:
\beqa\label{72}
\fl\;\;\;\;\;\Bvec^{(3)}_{(0)}=\frac{2\,\mu_0}{21}\mct^{(3)}\vert\vert\nablav^2\delta(\rvec)
-\frac{\mu_0}{6}\widetilde{\latens}\vert\vert\nablav^2\delta(\rvec)+\frac{\mu_0}{6}\widetilde{\latens}\,\Delta\delta(\rvec)
+\frac{\mu_0}{9}\Nsft^{(2)}\vert\times\vert\nablav^2\delta(\rvec)\ .
\eeqa
Let us consider the extension  for an arbitrary $n$ of
\beqa\label{73}
\fl\;\;\;\;\;\;\;\Bvec^{(n)}(\rvec)=\frac{\mu_0(-1)^{n-1}}{4\pi\,n!}\nablav\left(\nablav^n\vert\vert\frac{\msft^{(n)}}{r}\right)
+\frac{\mu_0(-1)^{n-1}}{n!}\left(\nablav^{n-1}\delta(\rvec)\right)\vert\vert\msft^{(n)}\ .
\eeqa
The generalized equation \eref{62} is given by
\beqa\label{74}
\msf_{i_1\dots i_n}=\mlr_{i_1\dots i_n}+\frac{1}{n}\suml^{n-1}_{\la=1}\eps_{i_\la i_nq}\Nsf^{(\la)}_{(i_1\dots i_{n-1}q)} \ ,
\eeqa
where it is introduced the $(n-1)$-th order tensor $\Nsft^{(n-1)}$, partial symmetric in the first $n-2$ indices and with null contraction of the last index $i_{n-1}$ with any of the indices $i_q,\;q\,<\,n-1$:
\beqa\label{75}
\fl\;\;\;\;\Nsf_{i_1\dots i_{n-1}}=\eps_{i_{n-1}ps}\msf_{i_1\dots i_{n-2}ps}=\frac{n}{n+1}\int_{\dom}\rmd^3x\,x_{i_1}\dots x_{i_{n-2}}\left[\rvec\times\left(\rvec\times\jvec\right)\right]_{i_{n-1}}\ .
\eeqa
Inserting equation \eref{74} in the expression of $\Bvec^{(n)}(\rvec)$, we consider the different terms from equation \eref{73}:
\beqan
\fl\nablav\left(\nablav^n\vert\vert\frac{\msft^{(n)}}{r}\right)&=&\nablav\left(\nablav^n\vert\vert\frac{\mlrt^{(n)}}{r}\right)+\frac{1}{n}\evec_i \,\left(\d_i\d_{i_1}\dots \d_{i_n}\frac{1}{r}\right)\suml^{n-1}_{\la=1}\eps_{i_\la i_nq}\Nsf^{(\la)}_{(i_1\dots i_{n-1}\,q)}\\
\fl&=&\nablav\left(\nablav^n\vert\vert\frac{\mlrt^{(n)}}{r}\right)
\eeqan
since all the terms from the last sum contain a contraction of the type $\eps_{i_li_nq}\d_{i_l}\d_{i_n}$ with $1\le l\le n-1$. 
\beqan
\fl\left(\nablav^{n-1}\delta(\rvec)\right)\vert\vert\msft^{(n)}= \left(\nablav^{n-1}\delta(\rvec)\right)\vert\vert\mlrt^{(n)}
+\evec_i\,\frac{1}{n}\left(\d_{i_1}\dots \d_{i_{n-1}}\delta(\rvec)\right)\suml^{n-1}_{\la=1}\eps_{i_\la i\,q}\Nsf^{(\la)}_{(i_1\dots i_{n-1}\,q)}\\
\fl=\left(\nablav^{n-1}\delta(\rvec)\right)\vert\vert\mlrt^{(n)}+\frac{n-1}{n}\evec_i\,\Nsf_{i_2\dots i_{n-1}\,q}\eps_{iqi_1}\d_{i_1}\d_{i_2}\dots i_{n-1}\delta(\rvec)\\
\fl=\;\mlrt^{(n)}\vert\vert\nablav^{n-1}\delta(\rvec)+\frac{n-1}{n}\Nsft^{(n-1)}\vert\times\vert\nablav^{n-1}\delta(\rvec) \ ,
\eeqan
using the notation \eref{65}. The final result for the substitution of equation \eref{74} in equation \eref{73} can be written as 
\beqa\label{76}
 \Bvec^{(n)}(\rvec)&=&\frac{\mu_0(-1)^{n-1}}{4\pi\,n!}\nablav\left(\nablav^n\vert\vert\frac{\mlrt^{(n)}}{r}\right)
+\frac{\mu_0(-1)^{n-1}}{n!}\mlrt^{(n)}\vert\vert\nablav^{n-1}\delta(\rvec)\nonumber\\
&+&\frac{\mu_0(-1)^{n-1}\,(n-1)}{n!\,n}\;\Nsft^{(n-1)}\vert\times\vert\nablav^{n-1}\delta(\rvec)\ .
\eeqa
Introducing the {\bf STF} tensor $\mct^{(n)}$ with the components given by an equation of the type \eref{34}
\beqan
\mlrt_{i_1\dots i_n}=\mct_{i_1\dots i_n}+\delta_{\{i_1i_2}\widetilde{\lasf}_{i_3\dots i_n\}}
\eeqan
and considering the different terms from equation \eref{76}, we obtain
\beqan
\nablav\left(\nabla^n\vert\vert\frac{\mlrt^{(n)}}{r}\right)=\nablav\left(\nabla^n\vert\vert\frac{\mct^{(n)}}{r}\right)
+\evec_i\d_i\left(\d_{i_1}\dots \d_{i_n}\frac{1}{r}\right)\,\delta_{\{i_1i_2}\widetilde{\lasf}_{i_3\dots i_n\}}\\
=\nablav\left(\nablav^n\vert\vert\frac{\mct^{(n)}}{r}\right)+\frac{n(n-1)}{2}\evec_i\left(\d_{i_3}\dots \d_{i_n}\,\Delta\frac{1}{r}\right)\widetilde{\lasf}_{i_3\dots i_n}\\
=\nablav\left(\nablav^n\vert\vert\frac{\mct^{(n)}}{r}\right)-2\pi\,n(n-1)\;\widetilde{\latens}^{(n-2)}\vert\vert\nablav^{n-1}\delta(\rvec)\ .
\eeqan
The second tensorial contraction from equation \eref{76} can be written as
\beqan
\mlrt^{(n)}\vert\vert\nablav^{n-1}\delta(\rvec)=\mct^{(n)}\vert\vert\nablav^{n-1}\delta(\rvec)+\evec_{i_n}\,\d_{i_1}\dots \d_{i_{n-1}}\,\delta(\rvec)\,\delta_{\{i_1i_2}\widetilde{\lasf}_{i_3\dots i_n\}}\ .
\eeqan
In the last expression, there are $n-1$ terms containing the factor $\delta_{i_qi_n},\;q=1,\dots n-1$ and $C^2_{n-1}=(n-1)(n-2)/2$ terms containing the factor $\delta_{i_qi_p}$ with $q$ and $s$ between $1$ and $n-1$, such that the last equation can be written as 
\beqa\label{77}
\mlrt^{(n)}\vert\vert\nablav^{n-1}\delta(\rvec)&=&\mct^{(n)}\vert\vert\nablav^{n-1}\delta(\rvec)+(n-1)\widetilde{\latens}^{(n-2)}\vert\vert\nablav^{n-1}\delta(\rvec)\nonumber\\
&&+\frac{(n-1)(n-2)}{2}\widetilde{\latens}^{(n-2}\vert\vert\nablav^{n-3}\Delta\delta(\rvec)\ .
\eeqa
Collecting all the above results, equation \eref{76} can be written as
\beqa\label{77p}
\fl\Bvec^{(n)}(\rvec)&=&\frac{(-1)^{n-1}\mu_0}{n!}\left[\frac{1}{4\pi}\nablav^{n+1}\vert\vert\frac{\mct^{(n)}}{r}+\mct^{(n)}\vert\vert\nablav^{n-1}\delta(\rvec)\right.\nonumber\\
\fl&&-\left.\frac{(n-1)(n-2)}{2}\widetilde{\latens}^{(n-2)}\vert\vert\nablav^{n-1}\delta(\rvec)
+\frac{(n-1)(n-2)}{2}\widetilde{\latens}^{(n-2)}\vert\vert\nablav^{(n-3)}\,\Delta\delta(\rvec)\right.\nonumber\\
\fl&&+\left.\frac{n-1}{n}\Nsft^{(n-1)}\vert\times\vert\nablav^{n-1}\delta(\rvec)\right]\ .
\eeqa
Employing equation \eref{71}, we finally write the singular $\delta$-form part of $\Bvec^{(n)}$:
\beqa\label{78}
\fl\Bvec^{(n)}(\rvec)&=&\frac{(-1)^{n-1}\mu_0}{n!}\left[-\frac{n}{2n+1}\mct^{(n)}\vert\vert\nablav^{n-1}\delta(\rvec)
-\frac{(n-1)(n-2)}{2}\widetilde{\latens}^{(n-2)}\vert\vert\nablav^{n-1}\delta(\rvec)\right.\nonumber\\
\fl&+&\left.\frac{(n-1)(n-2)}{2}\widetilde{\latens}^{(n-2)}\vert\vert\nablav^{(n-3)}\,\Delta\delta(\rvec)
+\frac{n-1}{n}\Nsft^{(n-1)}\vert\times\vert\nablav^{n-1}\delta(\rvec)\right]\ .
\eeqa
One can easily verify  that equations \eref{60} and \eref{72} are particular cases of equation \eref{78}.

\section{Conclusion}
The results of the present paper concerning the $\delta$-form singularities of the electromagnetic field in the static cases are not so appealing as the ones done in Ref. \cite{cvrz-arx}. However, in \cite{cvrz-arx} the corresponding results are obtained, in our opinion, without employing the full content of the multipole expansions. The unpleasant presence of the parameters $\latens$ and $\Nsft$ in the expressions of the $\delta$-type singularities is a consequence of the hypothesis that the basic multipolar moments are the primitive ones ($\psft^{(n)}$ in the electric case and $\msft^{(n)}$ in the magnetic one). The employment of irreducible representations by the {\bf STF} tensors has the advantage of simplicity in expressing some quantities in several circumstances. It seems, for example, that some trouble appears when employing the multipole expansions in spherical coordinates. Let us consider the  example of 
  the electrostatic potential multipole expansion:
\beqan
\Phi(\rvec)=\frac{1}{4\pi\eps_0}\suml^\infty_{l=0}\suml^{l}_{m=-l}\frac{Q_{lm}}{r^{l+1}}\,Y_{lm}(\theta,\,\varphi)
\eeqan
and the particular case $l=2$:
\beqa\label{c1}
\Phi^{(2)}(\rvec)=\frac{1}{4\pi\eps_0}\;\frac{1}{ r^3}\suml^2_{m=-2}Q_{2m}\,Y_{2m}\ .
\eeqa
The spherical moments $Q_{2m}$ are linear combinations of the components $\pc_{ij}$ of the {\bf STF} moment $\pct^{(2)}$. Writing this term in Cartesian coordinates,
\beqan
\Phi^{(2)}(\rvec)=\frac{1}{8\pi\eps_0}\psft^{(2)}\vert\vert\nablav^2\frac{1}{r}=
\frac{1}{8\pi\eps_0}\psf_{ij}\d_i\d_j\frac{1}{r}\ ,
\eeqan
and introducing the {\bf STF} tensor $\pc^{(2)}$,
\beqan
\Phi^{(2)}(\rvec)=\frac{1}{8\pi\eps_0}\left[\pc_{ij}\d_i\d_j\frac{1}{r}+\lasf \,\Delta\frac{1}{r}\right]\ ,
\eeqan
i.e
\beqan
\Phi^{(2)}(\rvec)=\frac{1}{8\pi\eps_0}\pc_{ij}\d_i\d_j\frac{1}{r}-\frac{1}{2\eps_0}\lasf\,\delta(\rvec)\ .
\eeqan
For the first term, equation \eref{16} gives the $\delta$-singularity 
\beqan
-\frac{1}{6\eps_0}\pc_{ij}\,\delta_{ij}\,\delta(\rvec)=0\ ,
\eeqan
since $\pc_{ii}=0$. Therefore,the electric dipolar  potential $\Phi^{(2)}$ has a delta-type singularity
\beqa\label{c2}
\left(\Phi^{(2)}\right)_{(0)}=-\frac{1}{2\eps_0}\latens\,\delta(\rvec).
\eeqa
But, for the term expressed in spherical coordinates (equation \eref{c1}), since there are no derivatives of $1/r$, one has no $\delta$-type singularities. 
\par In our opinion, equation \eref{c2} represents the correct result. From the given example we also see that some care is necessary when the invariance properties of the multipole expansions are used.
\par If correct, the results of the present paper can be useful in classical and quantum physics,  in the second case  starting with the problem of the hyperfine atomic structure.
\par The dynamic case will be treated similarly elsewhere.

\end{document}